\documentclass[pra,showpacs,twocolumn]{revtex4-1}

\usepackage{amsfonts,amssymb, amsmath,graphicx,mathrsfs,bm,appendix}
\usepackage{float}
\usepackage{txfonts}
\usepackage{subfigure}
\begin{document}

\title{Suppression of space broadening of exciton polariton beams by Bloch oscillation effects}

\author{Xudong Duan}
\author{Bingsuo Zou}
\author{Yongyou Zhang}
\email[Author to whom correspondence should be addressed. Electronic mail: ]{yyzhang@bit.edu.cn}
\affiliation{Beijing Key Lab of Nanophotonics $\&$ Ultrafine Optoelectronic Systems and School of Physics, Beijing Institute of Technology, Beijing 100081, China}%

\begin{abstract}
We theoretically study the transport of exciton polaritons under different applied photon potentials. The relation between the photon potentials and the thickness of the cavity layer is calculated by the finite element simulation. The theoretical analysis and numerical calculation indicate that the cavity photon potential is proportional to the thickness of the cavity layer with the coefficient being about $1.8$ meV/nm. Further, the periodic and linear photon potentials are considered to control the transport of the exciton polaritons in weak- and strong-field pump situations. In both situations the periodic potential cannot by itself effectively suppress the scatterings of the disorder potentials of the cavity photons and excitons and the nonlinear exciton-exciton interaction. When the linear potential is added to the cavity photons, the polariton transport exhibits the Bloch oscillation behavior. Importantly, the polariton Bloch oscillation can strongly suppress the space broadening due to the disorder potentials and nonlinear exciton-exciton interaction, which is beneficial for designing the polariton circuits.
\end{abstract}

\pacs{42.50.Ct, 42.79.Gn, 71.35.-y}
\maketitle

\section{Introduction}

Studies on polaritons, especially on microcavity polaritons (MPs), are developing rapidly in recent decades, though they were first predicted by Hopfield \cite{Hopfield1958Theory} and Agranoovich \cite{Agranovich1959} in the late 1950s. Pictorially, the MPs are a kind of photon dressed excitations: a quantum state with part of the cavity photon and part of the quantum well (QW) exciton. The strong coupling between the QW excitons and cavity photons takes overall responsibility for the form of the MPs. For the MPs, there is a sharp dip at the center of the dispersion which implies that the effective mass of the MPs is very light, and thus it is easy to achieve the Bose-Einstein condensation \cite{Kasprzak2006Bose, Byrnes2014Exciton}, superfluid \cite{Amo2009Superfluidity}, and vortex \cite{Rubo2007Half, Lagoudakis2009Observation}, for the MPs. In addition, the exciton part in the MPs leads to the nonlinear polariton-polariton interaction \cite{Zoubi2007Microscopic} and the numerous dynamics of the MPs \cite{Winkler2013Electroluminescence, Nakayama2014Photoluminescence, Yulin2008Dark, Karr2004Twin}. With the help of the dip dispersion and nonlinear interaction, people achieved or designend optical amplifiers \cite{Huynh2003Polariton} and switches \cite{Zhang2007Phase} based on the microcavities. These works have made the study on the MPs rapidly move forward.

With the developing of the nano-/micro-fabricating technologies, physical experimentalists manufactured many kinds of semiconductor microcavities (SMCs), including photonic dots \cite{Verger2006Polariton}, microcavity array \cite{Park200540000pixel}, one-dimensional microcavities \cite{Gamouras2013Energy-selective}. In these microcavities,  scalar potentials can be generated for the QW excitons or/and the cavity photons. For example, the polariton trap used in the Bose-Einstein condensation is induced by applying a mechanical stress \cite{Balili2007Bose-Einstein}. On the other hand, one can turn to the dependence of the cavity mode energy on the thickness of the cavity layer. With this idea, position-dependent potentials for the cavity photons can be designed into arbitrary geometries. Another method to introduce the photonic potential is to deposit surface optical structures on top of the SMC structure, such as metallic strips \cite{Lai2007Coherent}. These developments are the foundation of the accurate design of the microcavity devices, such as optical diodes \cite{Espinosa-Ortega2013Optical} and quantum circuits \cite{Probst2014Hybrid, Liew2008Optical, Espinosa-Ortega2013Complete}. Generally speaking, it is easier to design the photonic potentials than to design the excitonic ones. When integrating the polariton devices together, the communication among them also needs an accurate design on the microcavities to suppress the MP scatterings due to the system disorder potentials and the nonlinear exciton-exciton interaction.

\begin{figure}[htb]
  \centering
  \includegraphics[width=8.5cm]{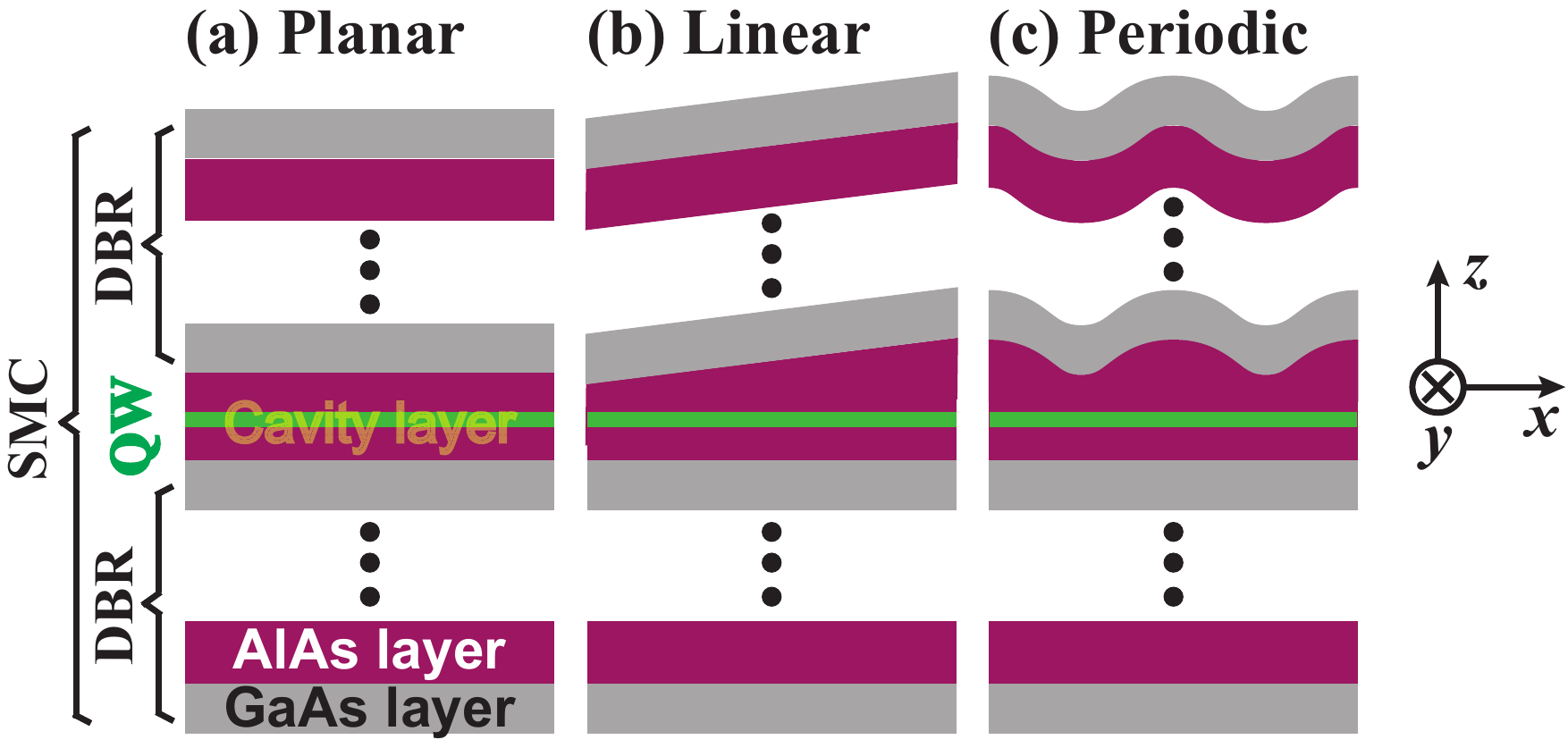}
 \caption{(Color online)  Schematics of the semiconductor microcavities with different geometries of the top DBR: (a) planar, (b) linearly tilted, and (c) periodic. The bottom panel gives the coordinate system adopted in this work.}
\end{figure}

The foundation of the communication among the polariton devices integrated together is to achieve the transport of the MPs without space broadening. This inspires us to study the influence of the applied potentials to the transport of the MPs. Because the photonic potentials are more convenient to achieve than the excitonic ones, we only consider the effects of the photonic potentials here. The photonic potentials originate from the defects of the microcavities or the geometry of the distributed Bragg reflectors (DBRs), referred to the Fig.~1. In the present work, we focus on three kinds of DBR geometries, namely, planar [in Fig.~1(a)], linearly tilted [in Fig.~1(b)], and periodic [in Fig.~1(c)]. For the planar SMCs, there is no applied photonic potential except a random photonic potential due to the random defects in the microcavities. Note that different geometries of the DBRs give different photonic potentials. Thus, we will determine their relations at first in the following section.

This work is organized as follows. In Sec.~II, we first find the relation between the photonic mode and the DBR geometry, and then determine the corresponding parameters of the photonic mode. Basing on these, we use the Gross-Pitaveskii equations to describe the dynamics of the MPs. In Sec.~III, we discuss the influence of the three geometries of the DBRs on the transport of the MPs. The Bloch oscillation could be observed under certain DBR geometry. At last, a brief conclusion is summarized in Sec.~IV.

\section{Theoretical model}

When the SMC is not planar, as shown in Figs.~1(b) to 1(c), the photon energy will depend on the transversal coordinates, which will influence the transport behavior of the MPs strongly. Without loss of generality, we assume the geometry of the DBR only depends on the coordinate $x$ and is independent of $y$, referred to Fig.~1. The start point of the present subsection is to find the dependence of the Hamiltonian of the cavity photons on the geometry of the DBR.

\subsection{Photonic Hamiltonian}

For a planar SMC shown in Fig.~1(a), the Hamiltonian of the cavity photons is well known, namely,
\begin{align}
H_c= E^c_0-{\hbar^2\over 2m_c}\nabla^2
\end{align}
where $\nabla^2={\partial^2\over\partial x^2}+{\partial^2\over\partial y^2}$. $E^c_0$ and $m_c$ are the fundamental energy and effective mass of the cavity photons, respectively. They can be roughly estimated by $E^c_0=\hbar c k_z/n_c$ and $m_c=n_c\hbar k_z/c=E^c_0n_c^2/c^2$. $\hbar$, $c$ and $n_c$ are the Plank constant, vacuum light speed, and  refract index of the cavity layer, respectively, and $k_z=N\pi/L_c$ with $L_c$ being the effective thickness of the cavity layer. The precondition of Eq.~(1) is that the in-plane wave vector of the cavity photon $k$ must be far less than $k_z$ which, for general microcavities, is satisfied.

When the effective thickness of the cavity is added by a small value $\Delta L_c$, the equation (1) can be cast into
\begin{align}
E^c(k)\approx (E^c_0+\Delta E_0^c)+\left(1-{\Delta E_0^c\over E_0^c}\right){\hbar^2 k^2\over 2m_c}
\end{align}
where $\Delta E_0^c$ represents the increase value of the fundamental photon energy due to $\Delta L_c$. For the GaAs-based SMCs, $E_0^c$ is about 1.5 eV, and the kinetic energy of the cavity photon is about several or tens of meV. Hence, the variation of the kinetic energy of the cavity photon due to the cavity thickness could be neglected, compared with the one of the fundamental energy. Therefore, we can rewrite the Hamiltonian of the cavity photon as follow,
\begin{align}
H_{ph}=-{\hbar^2\over 2m_c}\nabla^2+V_c(x)+E^c_0
\end{align}
where $E^c_0={\hbar c\over n_c} {N\pi\over L_0}$ with $L_0$ being a reference thickness of the cavity layer. $V_c(x)$ represents the photonic potential which could be estimated by
\begin{align}
V_c(x)=E_0^c\left[{L_0\over L_0+d(x)}-1\right]\approx -E_0^c{d(x)\over L_0}
\end{align}
where $d(x)$ is the thickness deviation of the cavity layer from $L_0$ and commonly, far less than $L_0$. The minus sign of the last term in Eq.~(4) indicates that the eigenfrequency of the cavity photon decreases with increasing the cavity thickness. The effectiveness of the Hamiltonian in Eq.~(3) can be confirmed by comparing its results with the numerical simulation of the finite-element method.

\begin{figure}[htb]
  \centering
  \includegraphics[width=8.5cm]{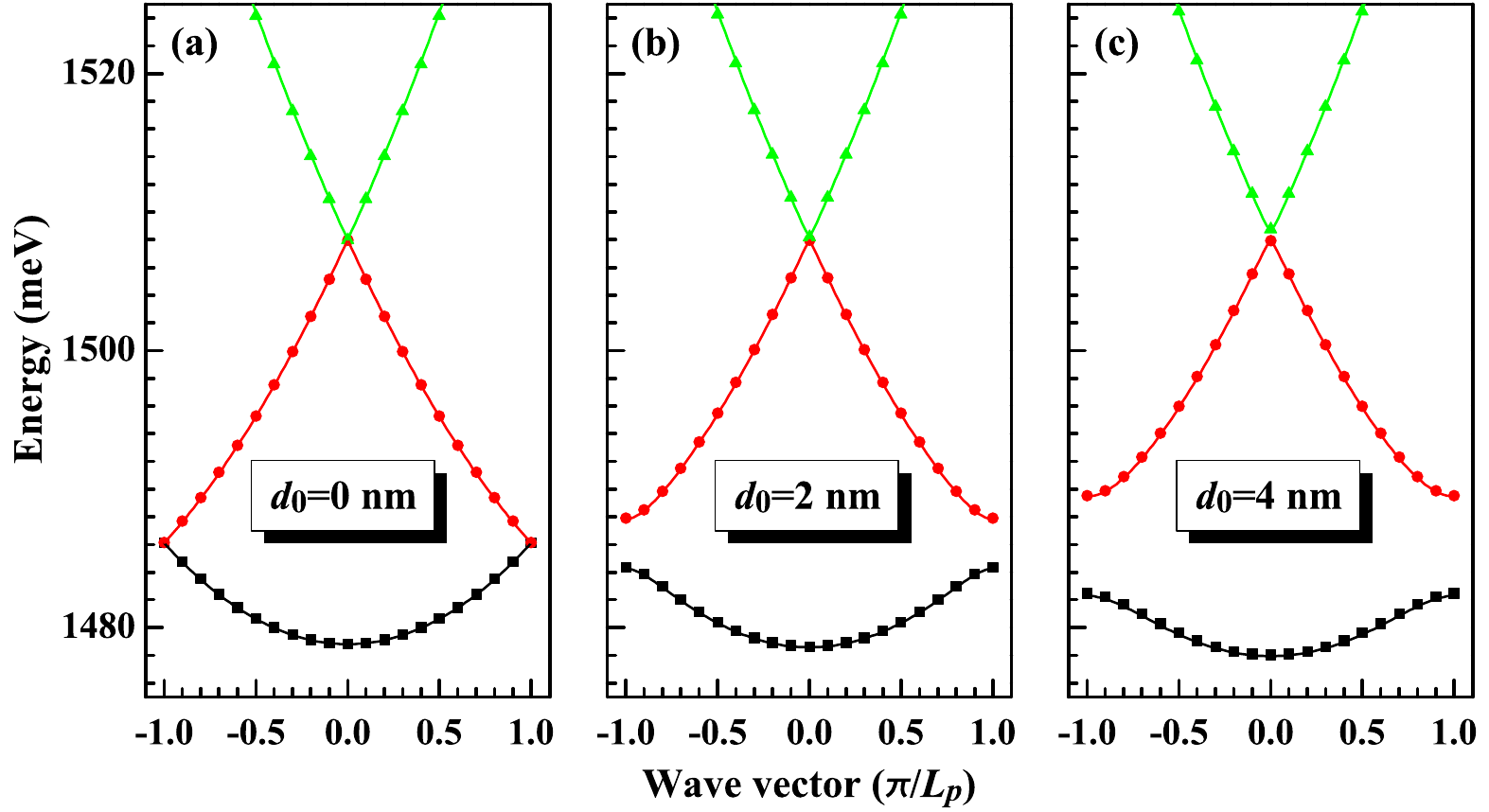}
 \caption{(Color online)  Photonic bands of the transversal electric modes in the periodic GaAs-based SMCs whose structures are shown in Fig.~1(c). The upper and lower DBRs hold 24 periodic cells. The cavity thickness of the SMC is described by $d(x)=d_0\cos(2\pi x/L_p)$ with $L_p=1.3~ \mu{\rm m}$ for all panels. The amplitude $d_0$ is given in each panel. All the data denoted by the dots are calculated from the finite element simulation, while all the solid lines are calculated from the Hamiltonian given in Eq.~(3).}
\end{figure}

We take the periodic GaAs-based SMC as an example, referred to Fig.~1(c). The refraction indexes of GaAs and AlAs are $n_a=3.6$ and $n_b=3.0$, respectively. The SMC is designed with the fundamental photon energy $E^c_0=1478.8$ meV. In addition, the potential $V_c(x)$ of the cavity photons are taken as the following form,
\begin{align}
V_c(x)&=-\alpha d(x)=-\alpha d_0\cos\left({2\pi\over L_p}x\right)
\end{align}
where $d_0$ and $L_p$ denote the amplitude and period of the variation of the cavity thickness. $\alpha$ is a coefficient that describes the photon potential owing to the variation of the cavity thickness.  With $d(x)$, one can numerically calculate the photonic bands of the SMC by the finite-element method. It is natural for that the band gaps at $\pm \pi/L_p$ increase with increasing $d_0$, just as shown in Figs.~2(a) to 2(c). In Fig.~2, all the data denoted by the dots are calculated from the finite element simulation, while all the solid lines are calculated from the Hamiltonian given in Eq.~(3).

In order to make the theoretical results meet the simulation data, we fit the dispersion in Fig.~2(a) by a parabolic dispersion (whose band gap is zero due to $d_0=0$) and find the effective mass of the cavity photon $m_c=3.1\times10^{-5}m_e$ where $m_e$ is the mass of a free electron. Using this effective mass of the cavity photon, we find the optimum value of $\alpha$ is about $\alpha=1.8 ~{\rm meV/nm}$ for the photon bands shown in Fig.~2(b). By Fig.~2(c), we further confirm the effectiveness of these two parameters. In practice, we also verify the availability of these two values of $m_c$ and $\alpha$ to the cases with other $d_0$. The calculations indicate that these two values are effective when $d_0\lesssim12$ nm. As a result, we can expect that the Hamiltonian in Eq.~(3) is available when the deviation of the cavity thickness is small. In the present work, this assumption will always be holden. Besides, we neglect the polarization difference between the transverse electrical mode and transverse magnetic mode for convenience. This means that the effective masses of the transverse electric mode and transverse magnetic mode are taken as equal to each other.

\subsection{Polaritonic crystal}

With the help of Eq.~(3) the polariton Hamiltonian can be written as
\begin{align}
H_{pl}\!=\!\!\int \!\!d^2\bm r \hat{\psi}^\dag(\bm r) { \hat {\bm h}}_0 \hat{\psi}(\bm r)+{1\over 2}g\!\!\int \!\!d^2\bm r\psi_x^\dag(\bm r)\psi_x^\dag(\bm r)\psi_x(\bm r)\psi_x(\bm r)
\end{align}
where $\hat{\psi}=\left(\hat{\psi}_c,~\hat{\psi}_x\right)^T$. $\hat{\psi}_c$ and $\hat{\psi}_x$ are the photonic and excitonic field operators, respectively. $g$ represents the nonlinear exciton-exciton coupling. The Hamiltonian $ \hat {\bm h}_0$ is
\begin{align}
\hat {\bm h}_0=\left(
\begin{array}{cc}
-{\hbar^2\over 2m_c}\nabla^2+V_c(x)+E^c_0&\Omega_R\\
\Omega_R & -{\hbar^2\over 2m_{x}}\nabla^2+E^x_0
\end{array}
\right)
\end{align}
where $m_x$ and $E_0^x$ are the exciton mass and fundamental energy, respectively. $\Omega_R$ is the Rabi coupling between the cavity photon and QW exciton. Note that for a non-planar SMC $\Omega_R$ shows a weak dependence on the coordinate. Such a weak dependence could be neglected when the deviation of the cavity thickness from its reference value is small, and we take $\Omega_R=2.65$ meV in this work. In this case, when $V_c(x)$ takes a periodic potential the Hamiltonian $ \hat {\bm h}_0$ leads to a polaritonic crystal whose band structure is shown in Fig.~3.

\begin{figure}[htb]
  \centering
  \includegraphics[width=8.5cm]{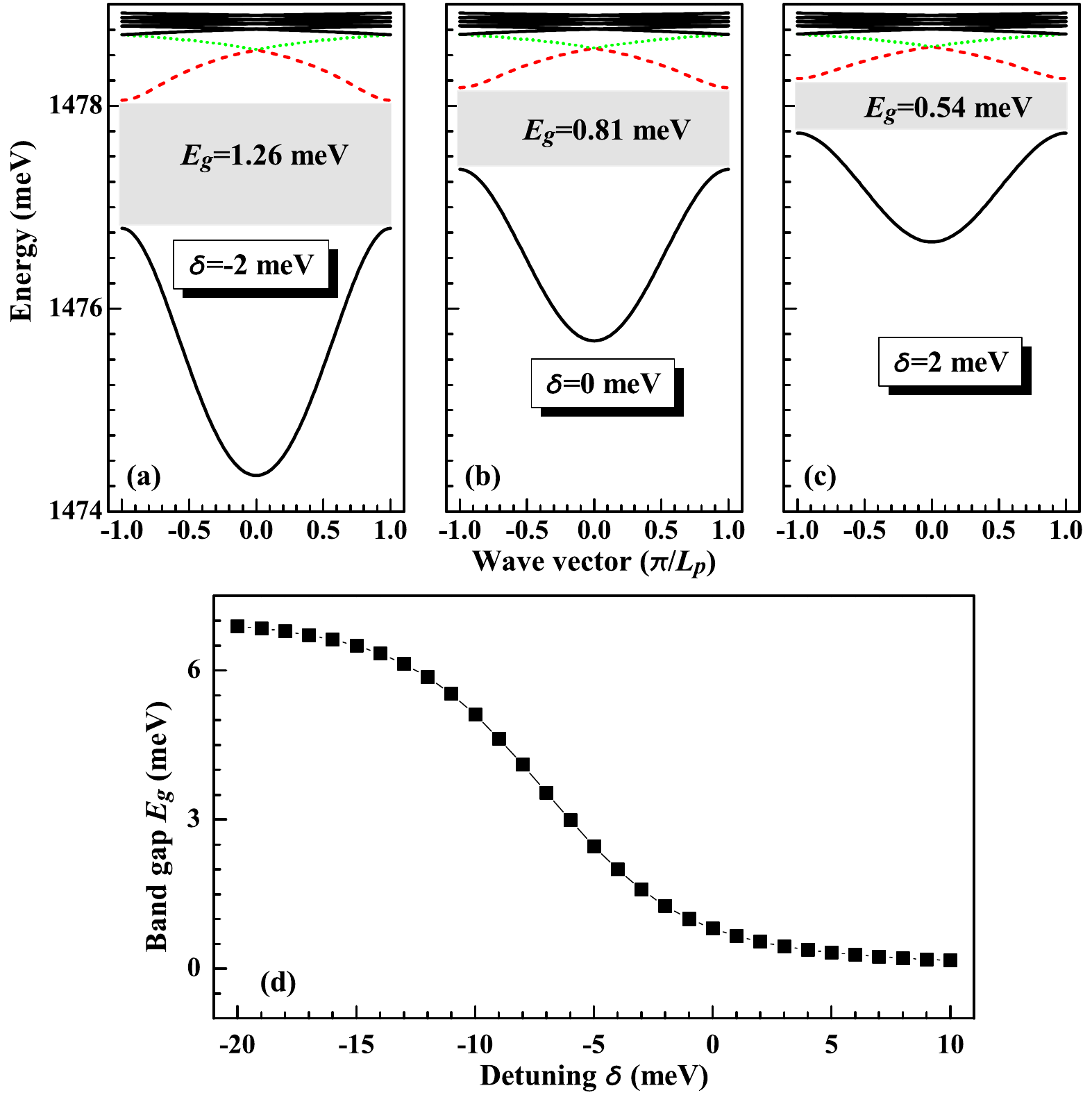}
 \caption{(Color online)  (a) to (c) Band structures of the polaritonic crystal under different photon-exciton detuning defined as $\delta=E_0^c-E_0^x$. The first band gaps are shown as the light gray regions in which the value of the band gap $E_g$ is shown. The top black dense lines represents the higher energy levels. Note that all energy bands shown in (a) to (c) belong to the lower polariton branch. (d) Variation of the first band gap $E_g$ with the photon-exciton detuning.
The SMC structure could refer to Fig.~1(c) and other parameters used in all panels are as follows: the upper and lower DBRs hold 24 periodic cells; the cavity thickness of the SMC is described by $d(x)=d_0\cos(2\pi x/L_p)$ with $L_p=1.3~ \mu{\rm m}$ and $d_0=4$ nm; and $E_0^x=1478.8$ meV, $m_x=0.22m_e$, and $\Omega_R=2.65$ meV.}
\end{figure}

Figure 3 shows the band structure of the polaritonic crystal under several different photon-exciton detuning $\delta$. The photon-exciton detuning is defined as $\delta=E_0^c-E_0^x$. In addition, the photonic potential $V_c(x)$ takes the form given in Eq.~(5). Note that only the lower-branch polariton bands are plotted and the lowest three bands are shown as the black solid, red dashed, and green dotted curves in Figs.~3(a), 3(b), and 3(c). Because of $m_x\gg m_c$, the exciton level is almost non-dispersive, and therefore, the polariton bands near the exciton level become denser and denser, displayed as the top dense black curves in Figs.~3(a), 3(b), and 3(c). For easy observation, the first forbidden band is denoted as the light gray region in which the value of the corresponding band gap $E_g$ is given. From Figs.~3(a) to 3(c), the band gap decreases as the photon-exciton detuning increases, referred to Fig.~3(d). This is due to that the exciton component in the lower polariton branch increases with increasing the detuning. Seen from Eq.~(7), the exciton potential is not introduced (which is not easy to achieve in the SMCs), thus the band gap of the polariton crystal shown in Fig.~3 originates from the photon component of the polaritons. When $\delta$ is much less than zero, the photon component dominates in the lowest several polariton bands, and so the polariton band gap approaches the photon one of 7.0 meV, referred to Fig.~2(c). When $\delta$ is much larger than zero, the dominant component in the lowest several polariton bands turns to be the excitons, and thus the polariton band gap turns to zero. Between these two situations, there is a region where the polariton band gap  varies with the detuning sharply, referred to Fig.~3(d).

\subsection{Gross-Pitaevskii equations}

In order to study the transport of the exciton polariton in the SMCs, such as polaritonic crystal, we follow a standard mean-field treatment on Eq.~(6), so that the evolution of the field functions can be described by the coupled Gross-Pitaevskii equations \cite{Zhang2009Boundary, Leyder2007Interference, Shelykh2006Polarization, Flayac2013An}
\begin{subequations}
\begin{align}
i\hbar{\partial\over\partial t}\psi_c&=\left[-{\hbar^2\over 2m_c}\nabla^2+V_c^D+V_c(x)+\delta-i{\gamma_c\over2}\right]\psi_c\nonumber\\
&\qquad\qquad\qquad\qquad+\Omega_R\psi_x+f_p(x,y)e^{-i\Delta t/\hbar},\\
i\hbar{\partial\over\partial t}\psi_x&=\left(-{\hbar^2\over 2m_x}\nabla^2+V_x^D-i{\gamma_x\over2}\right)\psi_x+\Omega_R\psi_c+g|\psi_x|^2\psi_x\
\end{align}
\end{subequations}
where $\hbar$ is the Planck constant which, for convenient, will be set to be 1 hereafter.
The exciton fundamental energy $E_0^x$ is chosen as the energy reference point. After the mean-field treatment, the photon and exciton fields, namely, $\psi_c$ and $\psi_x$, have been taken as complex numbers. $\gamma_c$ and $\gamma_x$ represent the photon and exciton losses, respectively. $\delta$ and $\Delta$ are the photon-exciton detuning and pump energy. $f_p$ measures the pump field whose space distribution and time dependence can be controlled by the applied pump lasers. $V_c^D$ and $V_x^D$􏰙 are taken as the static disordered potentials (DPs) for the cavity 􏰙photons and excitons, respectively􏰚. $V_c(x)$ represents the photon potential determined by the thickness of the cavity layer. With the help of the Gross-Pitaevskii equations, one can study the transport of the polaritons conveniently. What we should point out here is that the parameters on the photons, such as the photon mass $m_c$, cavity loss $\gamma_c$, and the photonic potential $V_c(x)$, are directly obtained from the finite element simulations as shown in Fig.~2. Thus, our theoretical results are in practice for the structures we considered.

For clarity, the parameters used throughout the present work are listed as follows. In the following calculations and discussions, we adopt the parameters of the typical GaAs-AlGaAs SMCs: $E_0^x=1478.8$ meV, $\delta=-2.0$ meV, $\gamma_c=\gamma_x=0.02$ meV, $m_c=3.1\times10^{-5}m_e$, $m_x=0.22m_e$, $\Omega_R=2.65$ meV, $g=0.015~{\rm meV\cdot\mu {\rm m}^2}$ \cite{Verger2006Polariton}, $\alpha=1.8~{\rm meV/nm}$. The static disordered potentials $V_c^D$ and $V_x^D$ 􏰙 are randomly generated, and their random-oscillation amplitudes are 0.4 meV and 0.1 meV, respectively.

\section{transport of exciton polaritons}

In this section, we study the transport of exciton polaritons in several different applied potentials. Besides the DPs of the QW excitons and cavity photons, we consider the two following photonic potentials,
\begin{subequations}
\begin{align}
V_c^P(x)&=V_P\times\cos\left({2\pi \over L_p}x\right),\\
V_c^L(x)&=V_L\times x.
\end{align}
\end{subequations}
Here, $V_c^P(x)$ and $V_c^L(x)$ are the periodic potential (PP) and linear potential (LP) of the cavity photons, respectively, and their strengths are set to be $V_P=7.2$ meV and $V_L=0.5$ meV$/\mu$m. The period of $V_c^P(x)$ is set to be $L_p=1.3~\mu$m. These two photonic potentials correspond to $d(x)=-4\cos\left({2\pi \over L_p}x\right)$ nm and $d(x)=-2.8\times10^{-4}x$ nm, respectively. In addition, the pump field takes the following form
\begin{align}
f_p(x,y)=F_pe^{-{x^2+y^2\over w^2}}e^{ik_py}
\end{align}
where $F_p$, $w$, and $k_p$ represent the amplitude, width, and wave vector along $y$ direction, respectively. In this section, $w$, and $k_p$ are taken to be $w=6~\mu$m, and $k_p=2~\mu{\rm}m^{-1}$. The pump frequency is set to be the energy of the lower-branch exciton polariton when the wave vector takes $k_p=2~\mu{\rm}m^{-1}$, namely,  $\Delta=-1.55$ meV. Thus, the pump scheme adopted in the present work is a resonant excitation.

For clearly displaying the transport of exciton polaritons, this section will be divided into two subsections according to the pump-field strength: 􏰂(A)􏰁 weak-field pump and (B) strong-field pump. For the former case, the nonlinear exciton-exciton interaction could be neglected, while for the later one, it must be considered.

\begin{figure}[htb]
  \centering
  \includegraphics[width=7.0 cm]{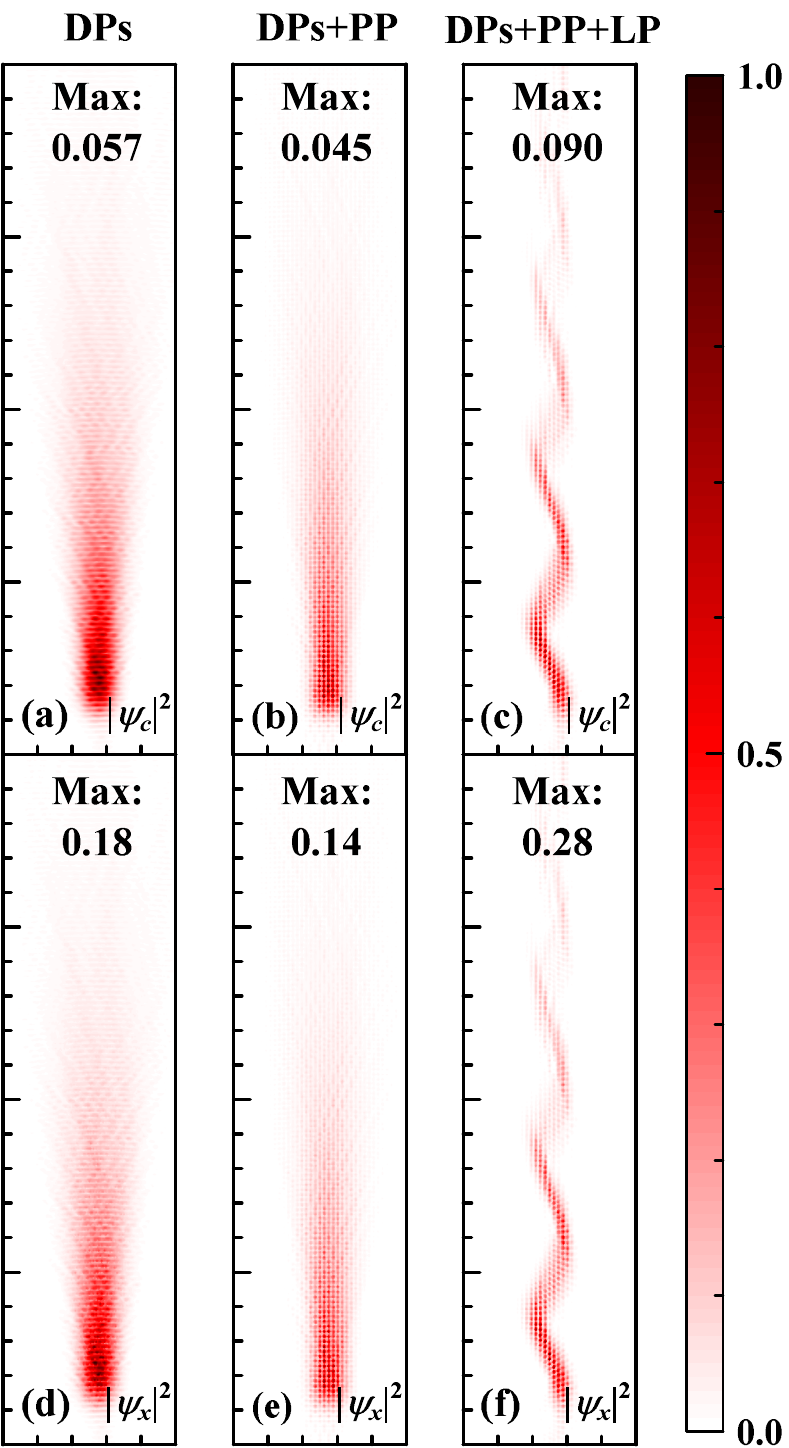}
 \caption{(Color online)  Distributions of photon fields $|\psi_c|^2$ (a-c) and their corresponding exciton fields $|\psi_x|^2$ (d-f) in a rectangular area of $50~\mu{\rm m}\times 200~\mu{\rm m}$ under weak-field pump. For easy observation, all field distributions are normalized to their maximum values, respectively, and their maximum values are shown in each panel. For each column, the adopted potentials are denoted on top: DPs, PP, and LP represent the disorder potentials of the cavity photons and excitons, cavity-photon periodic potential, and cavity-photon linear potential, respectively. Here the pump field is weak, and its amplitude is $F_p=0.1~{\rm meV}\cdot\mu{\rm m}^{-1}$.}
\end{figure}

\subsection{Weak-field pump}

Figure 4 shows the transport of the exciton polaritons under different applied photonic potentials which are denoted on top. When only the disorder potentials $V_c^D$ and $V_x^D$ are introduced (this case will be denoted as DPs hereafter), referred to Figs.~4(a) and 4(d), the exciton polaritons are scattered away from the transport direction, namely, direction of the vertical. Therefore, the polariton beam is broadened in its transport path. When a periodical potential for the cavity photons given in Eq.~(9a) is added (this case will be denoted as DPs+PP hereafter), this behavior can also be seen, referred to Figs.~4(b) and 4(e), but weakly suppressed. The beam broadening of the exciton polaritons can be further suppressed by adding a linear photon potential given in Eq.~(9b) [see Figs.~4(c) and 4(f)] (this case will be denoted as DPs+PP+LP hereafter). Obviously, the transport of the exciton polaritons exhibits an oscillation behavior which is the well-known Bloch oscillation (BO). In order to confirm the oscillation in Figs.~4(c) and 4(f) being the BO, we first neglect the optical pump field, disorder potentials in Eq.~(8), and then under slow wave approximation transform them into the following form
\begin{subequations}
\begin{align}
i\lambdabar\frac{\partial\varphi_c}{\partial y}&=\left(-\frac{\lambdabar^2}{2n}\frac{\partial^{2}}{\partial x^2}+{\cal V}_c(x)+\varepsilon_c\right)\varphi_{c}+\Omega_c\varphi_x,\\
i\lambdabar{\partial\varphi_x\over\partial y}&=\left(-\frac{\lambdabar^{2}}{2n}{\partial^2\over\partial x^2}+\varepsilon_x+g|\varphi_x|^2\right)\varphi_x+\Omega_x\varphi_{c},
\end{align}
\end{subequations}
where $\varphi_{c(x)}=\psi_{c(x)} e^{-ik_py+i\Delta t}$. Other parameters are defined as follows: $\lambdabar={\hbar^2k_p/ m_x}$, $n=\lambdabar k_p$, $\Omega_x=\Omega_R$, $\Omega_c={m_x\Omega_R/ m_c}$, $\varepsilon_x={\hbar^2k_p^2/ 2m_x}-\Delta$, $\varepsilon_c={m_c\over m_x}\left(\delta+{\hbar^2k_p^2/ 2m_c}- \Delta\right)$, and ${\cal V}_c(x)={m_c\over m_x}V_c(x)$. From Eq.~(11), the BO period $L_B$ can be found to be $L_B=2\pi \lambdabar m_x/(m_c V_LL_p)$ \cite{Dreisow2009Bloch-Zener, Longhi2006Optical, Zhang2010Bloch}$=48.3~\mu{\rm m}$ which is consistent with Figs.~4(c) and 4(f). Thus, the oscillations appeared in Figs.~4(c) and 4(f) correspond to the Bloch oscillation.

\begin{figure}[htb]
\centering
\includegraphics[width=8.6cm]{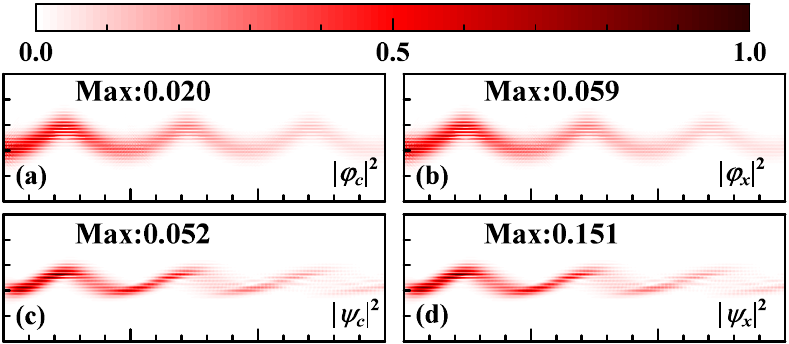}
\caption{(Color online)  Distributions of photon field  (a) and exciton field (b) calculated from Eq.~(11), and distributions of photon field (c) and exciton field (d) calculated from Eq.~(8). In calculating (a) and (b), the initial fields are taken as $\varphi_c(x)|_{y=0}=F_pe^{-x^2/w^2}$ and $\varphi_x(x)|_{y=0}=0$ with $F_p=0.1~{\rm meV}\cdot\mu{\rm m}^{-1}$. Other parameters for all panels are the same in calculation and consistent with the ones used in Figs.~4(c) and 4(f), except without the disorder potentials for the photons and excitons. All panels are $150~\mu{\rm m}\times50~\mu{\rm m}$.}
\end{figure}

We can further confirm this by comparing the distributions of the exciton and photon fields calculated from Eq.~(11) and Eq.~(8), respectively (see Fig.~5). Obviously, their field distributions hold almost the same transport patten, except the values and widths of the field beams. The differences of the field values and beam widths between them are due to their different excitations. In calculating Figs.~5(a) and 5(b) through Eq.~(11), the excitation is given by the initial fields of the photons and excitons, namely, $\varphi_c(x)|_{y=0}=F_pe^{-x^2/w^2}$ and $\varphi_x(x)|_{y=0}=0$ with $F_p=0.1~{\rm meV}\cdot\mu{\rm m}^{-1}$. In calculating Figs.~5(c) and 5(d) through Eq.~(8), the excitation  is given by Eq.~(10) with the same parameters with Fig.~4(c) and 4(f). This different excitation does not influence the conclusion that the oscillations in Figs.~4(c) and 4(f), as well as in Figs.~5(c) and 5(d), are the Bloch oscillation. The BO plays an important role in achieving the long-distance polariton transport with a narrow beam width. Comparing Figs.~4(c) with 4(a) and 4(b), we can find that the field beam width in Fig.~4(c) almost does not change with increasing the transport distance. In order to describe this effect, we introduce the effective width of the field beam, $W_{eff}$, as follow,
\begin{align}
W_{eff}=\left[\frac{\int \left(x-\bar{x}\right)^{2}|\psi_c|^{2}dx}{\int |\psi_c|^{2}dx}\right]^{1/ 2}
\end{align}
where $\bar{x}=\int x|\psi_c|^{2}dx / \int |\psi_c|^{2}dx$, being the center position of the field beam along direction of $x$.

\begin{figure}[htb]
\centering
\includegraphics[width=8.6cm]{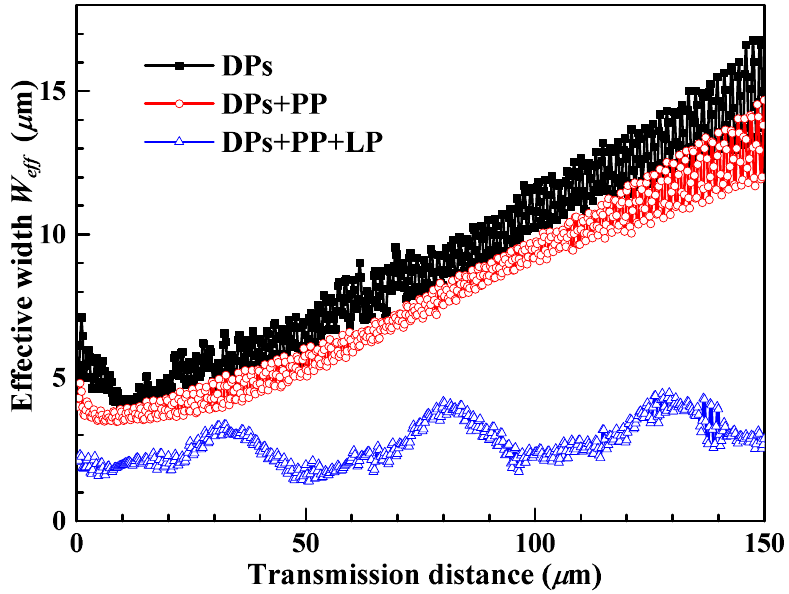}
\caption{(Color online)  Variation of the effective widths of the photon field with the transport distance under different applied potentials. The parameters used are the same to Fig.~4. For each curve, the adopted potentials are denoted: DPs, PP, and LP represent the disorder potentials of the cavity photons and excitons, cavity-photon periodic potential, and cavity-photon linear potential, respectively.}
\end{figure}

We plot the effective width $W_{eff}$ as the function of the transport distance in Fig.~6. Due to the scattering of the disorder potentials of the cavity photons and excitons, the photon field beams become wider and wider with increasing the transport distance for the cases of DPs and DPs+PP, referred to the upper two curves. For them, $W_{eff}$ approximately linearly increases with increasing the transport distance, which is mainly determined by the strength of the DPs. However, when the linear potential for the cavity photons is applied, the width of the photon field beam almost does not increase after a long transport distance and approximates the width of the pump field. We attribute this phenomenon to the Bloch oscillation which also leads to periodical oscillation of $W_{eff}$, as shown by the lowest curve in Fig.~6. The oscillation period of $W_{eff}$ is the same to the period of the BO, being about $48.5~\mu{\rm m}$. If only the LP is applied for the cavity photons, the photon and exciton field patterns look like the trace of the free-fall body until they meet the cavity boundary (which can reflect them with solid boundary conditions). This implies when only the LP is applied, the transport of the polaritons is not along one main direction, which is why we do not consider them in the present work. According to above discussions, we could get that the BO effects of the polariton can be used to achieve the transport without space broadening for the weak-field pump situation.

\begin{figure}[htb]
  \centering
  \includegraphics[width=7.0cm]{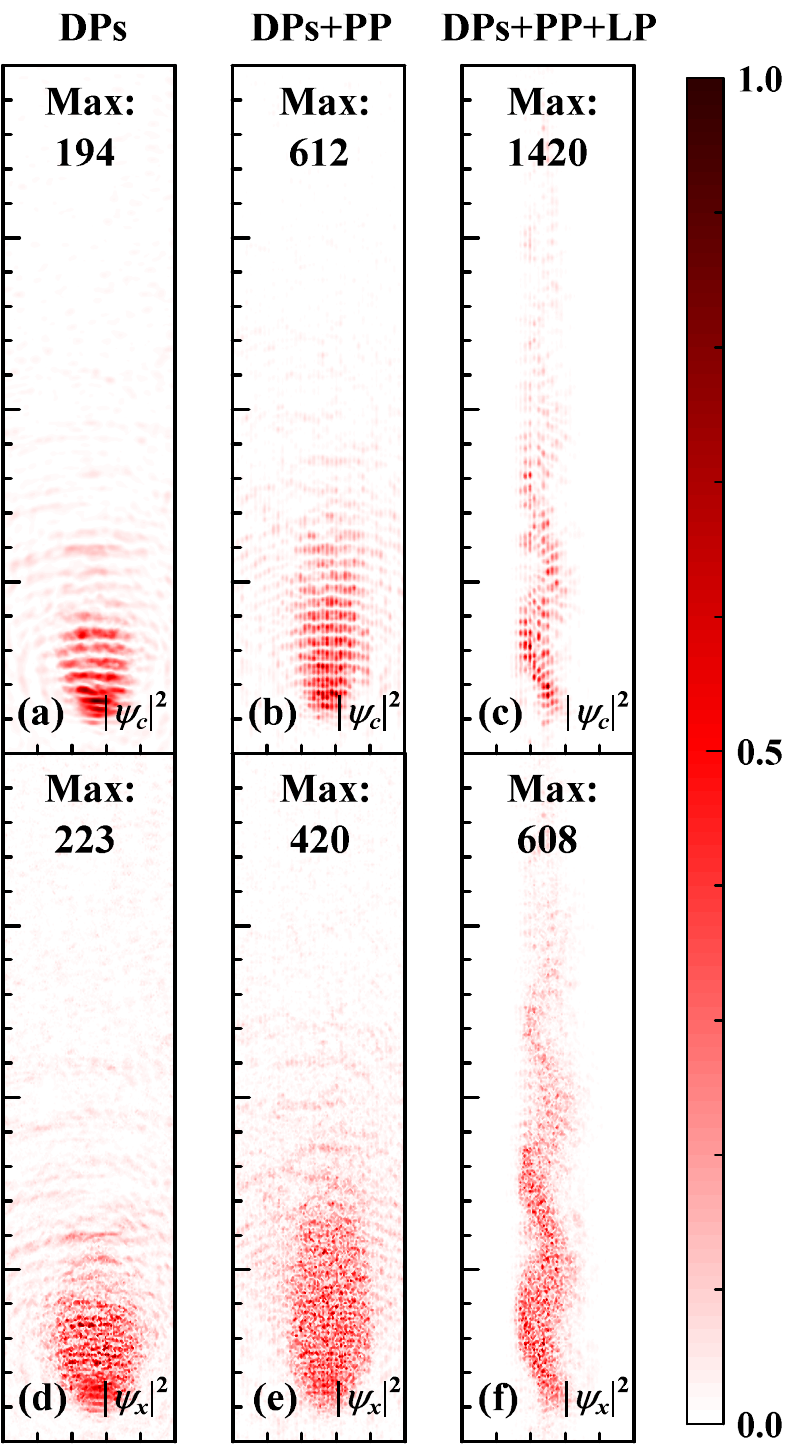}
 \caption{(Color online)  Distributions of photon fields $|\psi_c|^2$ (a-c) and their corresponding exciton fields $|\psi_x|^2$ (d-f) in a rectangular area of $50~\mu{\rm m}\times 200~\mu{\rm m}$ under strong-field pump. Other indications and parameters are the same with Fig.~4 except the amplitude of the pump field $F_p=10~{\rm meV}\cdot\mu{\rm m}^{-1}$.}
\end{figure}

\subsection{Strong-field pump}

In this subsection, we study the transport of the polaritons with the strong-field pump. For the situations of the weak-field pump, it is known that the nonlinear exciton-exciton interaction could be neglected, while this nonlinear interaction plays important roles in the situation of the strong-field pump. It can enhance the scattering of DPs during the polariton transport, as shown in Fig.~7. Comparing with Fig.~4, the field patterns become wider, especially for the cases of DPs and DPs+PP. Because of the strong nonlinear interaction, the BO patterns shown in Figs.~7(c) and 7(f) also become indistinct. Note that pump frequency equals the exciton energy, thus for the weak-field pump cases the excitation for the exciton field is resonant. Together with the $V_c^D$ amplitude being larger than the $V_x^D$ one, we can see that the exciton field is stronger than the photon one for the weak-field pump (see Fig.~4). However, for the strong-field pump the excitation for the exciton becomes non-resonant due to the strong influence of the exciton-exciton interaction. Thus, the exciton field approximates or becomes weaker than the photon one for the strong-field pump (See Fig.~7). Though the nonlinear exciton-exciton interaction exerts strong influence on the polariton transport, the BO still limits the space broadening of the polariton field, referred to Figs.~7(c) and 7(f). This effect is also confirmed by Fig.~8.

\begin{figure}[htb]
\centering
\includegraphics[width=8.6cm]{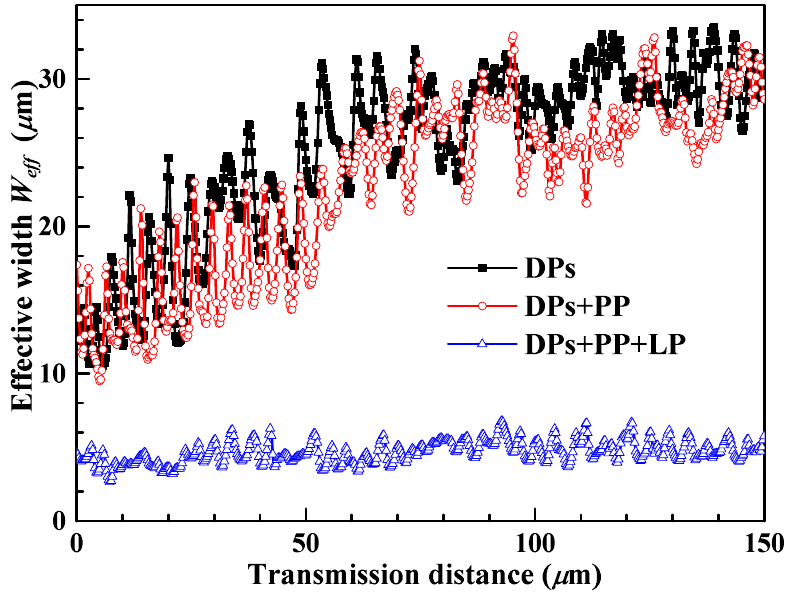}
\caption{(Color online)  Variation of the effective widths of the photon field with the transport distance under different applied potentials. The parameters used are the same to Fig.~7.}
\end{figure}

The BO makes the polariton fields move along one main direction without space broadening, as shown by the lowest curve in Fig.~8. The lowest curve in Fig.~8 indicates that the average of $W_{eff}$ is about 4.0 $\mu$m. This value of $W_{eff}$ approximates the width of the pump field $w$ for the case of DPs+PP+LP and does not increase with increasing with the transport distance, though it is a little larger than the situation of the weak-field pump. However, $W_{eff}$ increases with increasing the transport distance for the cases of DPs and DPs+PP, which is similar to the weak-field pump situations. Therefore, the BO not only can shield the scattering due to the DPs but also can shield the scattering due to the nonlinear exciton-exciton interaction. According to this, the BO can be used to achieve the polariton transport without the space broadening.

\section{conclusion}

We have theoretically investigated the transport of exciton polaritons under different applied photon potentials in the semiconductor microcavities. First, we proved that the cavity photon potential is proportional to the thickness deviation of the cavity layer from the reference value and the corresponding ratio coefficient is about $1.8$ meV/nm fitted from the finite element method. Then, we studied the polariton transport under DPs of the QW excitons and cavity photons. When the periodic and linear potentials are applied to the cavity photons, the Bloch oscillation of the polaritons can be observed. In the situation of the weak-pump field, the Bloch oscillation of the polaritons is further confirmed by the method of slow wave approximation. Subsequently, we found that the Bloch oscillation can effectively suppress the scattering of the disorder potentials and the nonlinear exciton-exciton interaction. The corresponding behavior is that the effective width of the photon and exciton fields does not increase with increasing the transport distance, no matter in the situations of the weak- or strong-field pumps. Thus, we conclude that the Bloch oscillation effects of the polaritons can be used to achieve the transport without space broadening, which is beneficial for designing the polariton circuits.

\section*{Acknowledgements}
This work is supported by NSFC (Grant No. 11304015), Beijing Higher Education Young Elite Teacher Project (Grant No. YETP1228), BIT Foundation for Basic Research (Grant Nos. 20121842005, 20131842002).

\bibliographystyle{aipnum4-1}
%

\end{document}